\documentclass[journal=jacsat,manuscript=article]{achemso}

\usepackage{epstopdf}
\usepackage{graphicx}
\usepackage{amstext}
\usepackage{amssymb}
\usepackage{amsbsy}
\usepackage{amsmath}
\usepackage{xcolor}
\usepackage{diagbox}
\usepackage{multirow}
\usepackage{array}

\newcolumntype{C}[1]{>{\centering\arraybackslash}p{#1}}
\usepackage[colorlinks,urlcolor=blue,linkcolor=blue,citecolor=blue]{hyperref}
\author{Hao-Yue Zhang}
\affiliation{School of Physics and Astronomy, Applied Optics Beijing Area Major Laboratory, Beijing Normal University, Beijing 100875, China}
\alsoaffiliation{Key Laboratory of Multiscale Spin Physics, Ministry of Education, Beijing Normal University, Beijing, China}

\author{Yi-Xuan Yao}
\affiliation{School of Physics and Astronomy, Applied Optics Beijing Area Major Laboratory, Beijing Normal University, Beijing 100875, China}
\alsoaffiliation{Key Laboratory of Multiscale Spin Physics, Ministry of Education, Beijing Normal University, Beijing, China}

\author{Bin-Yao Huang}
\affiliation{School of Physics and Astronomy, Applied Optics Beijing Area Major Laboratory, Beijing Normal University, Beijing 100875, China}
\alsoaffiliation{Key Laboratory of Multiscale Spin Physics, Ministry of Education, Beijing Normal University, Beijing, China}
\alsoaffiliation{School of Science, Chongqing University of Posts and Telecommunications, Chongqing 400065, China}

\author{Jing-Yi-Ran Jin}
\affiliation{School of Physics and Astronomy, Applied Optics Beijing Area Major Laboratory, Beijing Normal University, Beijing 100875, China}
\alsoaffiliation{Key Laboratory of Multiscale Spin Physics, Ministry of Education, Beijing Normal University, Beijing, China}

\author{Qing Ai}
\email{aiqing@bnu.edu.cn}
\affiliation{School of Physics and Astronomy, Applied Optics Beijing Area Major Laboratory, Beijing Normal University, Beijing 100875, China}
\alsoaffiliation{Key Laboratory of Multiscale Spin Physics, Ministry of Education, Beijing Normal University, Beijing, China}

\title{Non-Hermitian Hamiltonian Approach for Two-Dimensional Coherent Spectra of Driven Systems}

\abbreviations{2DCS,RF,NHH,RP,EIT}

\begin{document}

\begin{tocentry}
\includegraphics[width=8.25cm]{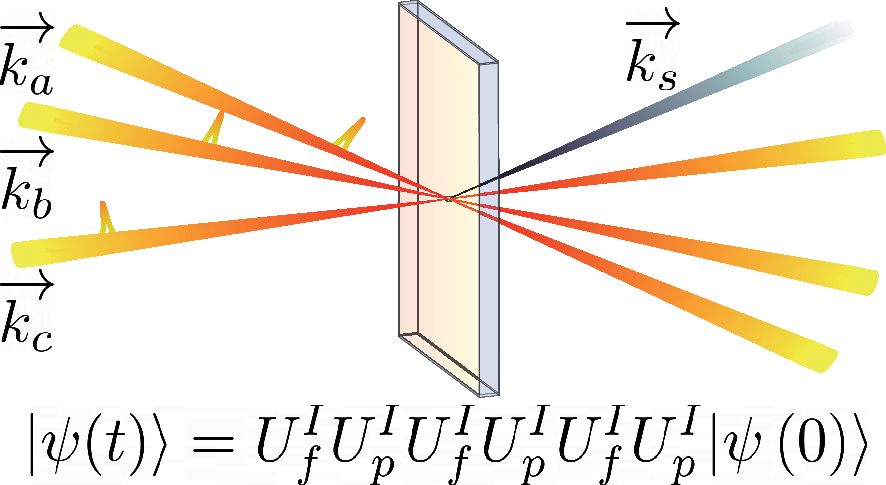}
\end{tocentry}

\begin{abstract}

Two-dimensional coherent spectroscopy (2DCS) offers significant advantages in terms of high temporal and frequency resolutions and signal-to-noise ratio. Until now, the response-function (RF) formalism has been the prevalent theoretical description. In this study, we compare the non-Hermitian Hamiltonian (NHH) method with the RF formalism in a three-level system with a constant control field. We obtain the signals from both approaches and compare their population dynamics and 2DCS. We propose the quasi-Green functions for the NHH method, which allows all dominant Liouville paths to be inferred. We further simulate the 2DCS of Rh(CO)$_2$C$_5$H$_7$O$_2$ (RDC) dissolved in hexane with the NHH method, which is in good agreement with the previous experiments. Although the NHH method overestimates relaxations, it provides all important paths by analytical solutions, which are different from the four paths used in the RF formalism. Our results demonstrate that the NHH method is more suitable than the RF formalism for investigating the systems including relaxation and control fields via the 2DCS.

\end{abstract}

\section{Introduction}
\label{sec:introduction}

Two-dimensional coherent spectroscopy (2DCS) is a technique that applies sequential pulses to materials or biological systems and receives output signals rich in structural and dynamic information \cite{Aue-1976,mukamel-2000,WOS:000352259800029,WOS:000461530000001,Sun2023AQT}. In 1998, Jonas et al. first experimentally demonstrated two-dimensional electronic spectroscopy \cite{HYBL1998307}. Subsequently, Asplund et al. demonstrated two-dimensional infrared spectroscopy \cite{Asplund2000}. This technology has rapidly advanced in both experimental methods and theoretical approaches since it was first proposed. Nowadays, this technique is implemented with a wide range of pulse wavelengths, from UV to microwave \cite{HYBL1998307,Myers2008,WOS:000270766800043,Kearns2017,Oliver2015,Courtney2015,Song2019,Duchi2021,Nardin2013}, providing temporal resolutions ranging from milliseconds to attoseconds \cite{WOS:000270766800043,WOS:000309566400010,WOS:000296494700045,WOS:000315525500002,WOS:000426497900006}. It is widely employed in physical chemistry, condensed-matter physics, and biophysics for exploring phenomena such as energy transfer \cite{Engel2007,lewis-2012,mehlenbacher-2016,WOS:000331947300005,WOS:000352259800031}, molecular vibrations \cite{golonzka-2001,butkus-2012}, electronic transitions and relaxations \cite{WOS:000269861400017,nagahara-2024,krecik-2015}, chemical reactions \cite{jansen-2007,herrera-2009}, as well as structures like proteins \cite{Tiwari2018,WOS:000408519200005,WOS:000723785200001}. The means to implement 2DCS experimentally are diverse \cite{bax-1986,kessler-1988,hamm-2011,hamm-2008,biswas-2022,fuller-2015}. The common principle involves the application of pulse sequences with different time intervals to samples, in which the pulses are well separated. The emitted polarization signals are projected onto two independent frequency axes, creating a map known as 2DCS.

On the other hand, one of the widely-used methods for the theoretical description of the 2DCS is the response-function (RF) formalism \cite{mukamel-1995,Deng2024PRA,WOS:000170647600002,Noblet2021}. This approach employs the double-sided Feynman diagrams, each corresponding to a Liouville path, to describe the polarization. The signals are viewed as the responses to the perturbations induced by the three applied pulses. However, the number of diagrams to be drawn and calculated may reach as many as hundreds if the system involves more external fields and energy levels. In such cases, the RF formalism may lead to the omission of some important diagrams. For example, an additional field is applied to induce transitions in the system. It may become exhausting to draw all possible paths to obtain the analytical solution, although most of them do not contribute significantly. On the other hand, the non-Hermitian Hamiltonian (NHH) method, which is widely-used in investigating PT symmetry \cite{Bender1998PRL,Ashida2017,Feng2017}, photosynthetic light harvesting \cite{Olaya-Castro2008PRB,Celardo2012,Nesterov2013}, chemical reaction and avian compass \cite{JONES201090,Haberkorn1976,maeda_chemical_2008}, and negative refraction in M\"{o}bius molecules \cite{Fang2016PRA}, may
 overcome these disadvantages. By omitting the quantum-jump terms, it obtains a reliable solution to the quantum master equation and thus achieves a balance between computational difficulty and analytical-solution-based description  \cite{WOS:000350274700003}. Nevertheless, working with non-Hermitian operators still requires careful consideration in interpreting the physical processes underlying the results.
Despite the RF formalism and the NHH method, other approaches with different advantages for simulating 2DCS include the hierarchical equations of motion approach \cite{tanimura-2020}, the nonperturbative response function formalism \cite{Chen-2017}, the numerical integration of the Schr\"odinger equation scheme \cite{sardjan-2020,Eric2023}, the non-linear exciton equations method \cite{mukamel-2000}, the non-linear exciton propagation method \cite{falvo-2009} and the full
 cumulant expansion formalism \cite{cupellini-2020}.

 In this article, we compare the NHH method and the RF formalism in a three-level system with a constantly-applied control field. It is a typical driven system, characterized by additional electromagnetic fields beyond the pulse sequence used in 2DCS. Such systems have been applied in Rb atoms and Cs$_{2}$ molecules to induce electromagnetically-induced transparency \cite{Jin2025,liu-2020}, and in 1,2-propanediol for enantiodetection \cite{Cai2022PRL}. The introduction of these fields facilitates the generation of novel phenomena, allowing for the combination with other techniques or the execution of more refined operations. The RF formalism is used as a reference because it is a widely-used method that can provide analytical solutions with a clear physical picture. Although Lemmer et al. first applied the NHH method to 2DCS \cite{WOS:000350274700003}, they did not focus on the approach itself but rather on its application to study ion Coulomb crystals. In contrast, our work concentrates on the similarities and differences between these two methods in calculating 2DCS. We validate their consistency by comparing the 2DCS when they follow identical Liouville paths. We discuss their respective advantages and disadvantages, as well as their scopes of application. Particularly, we propose the quasi-Green function for interpreting the 2DCS by the NHH method, which significantly simplifies the analytical challenges. Then we use the NHH method to simulate the 2DCS of Rh(CO)$_2$C$_5$H$_7$O$_2$ (RDC) dissolved in hexane to further validate this approach. Overall, we believe that the NHH method provides a reliable description for 2DCS, particularly in  driven or complex systems, and can overcome the limitations of the RF formalism in these scenarios.

We first illustrate the RF formalism and the NHH method, in which the quasi-Green function is proposed as a counterpart to the Green function in the RF formalism. We then present their double-sided Feynman diagrams and the analytical expressions of the rephasing (RP) signal. We next provide a comparison of the Green and quasi-Green functions, the population dynamics, the conservation of the population and the 2DCS by the two approaches, focusing mainly on their differences in coherent and incoherent behavior. The advantages and disadvantages, as well as the difference and consistency of the two methods are discussed. We further apply the NHH method to RDC dissolved in hexane and discussed the results in relation to its energy levels and the existing experimental results. Finally, we summarize our results and offer a perspective on the application of the NHH method.

\section{Theoretical approaches} \label{sec:Theoretical approaches}

\subsection{Response Function Formalism} \label{sec:Response Function Formalism}

\begin{figure}[h]
\includegraphics[width=8.45cm]{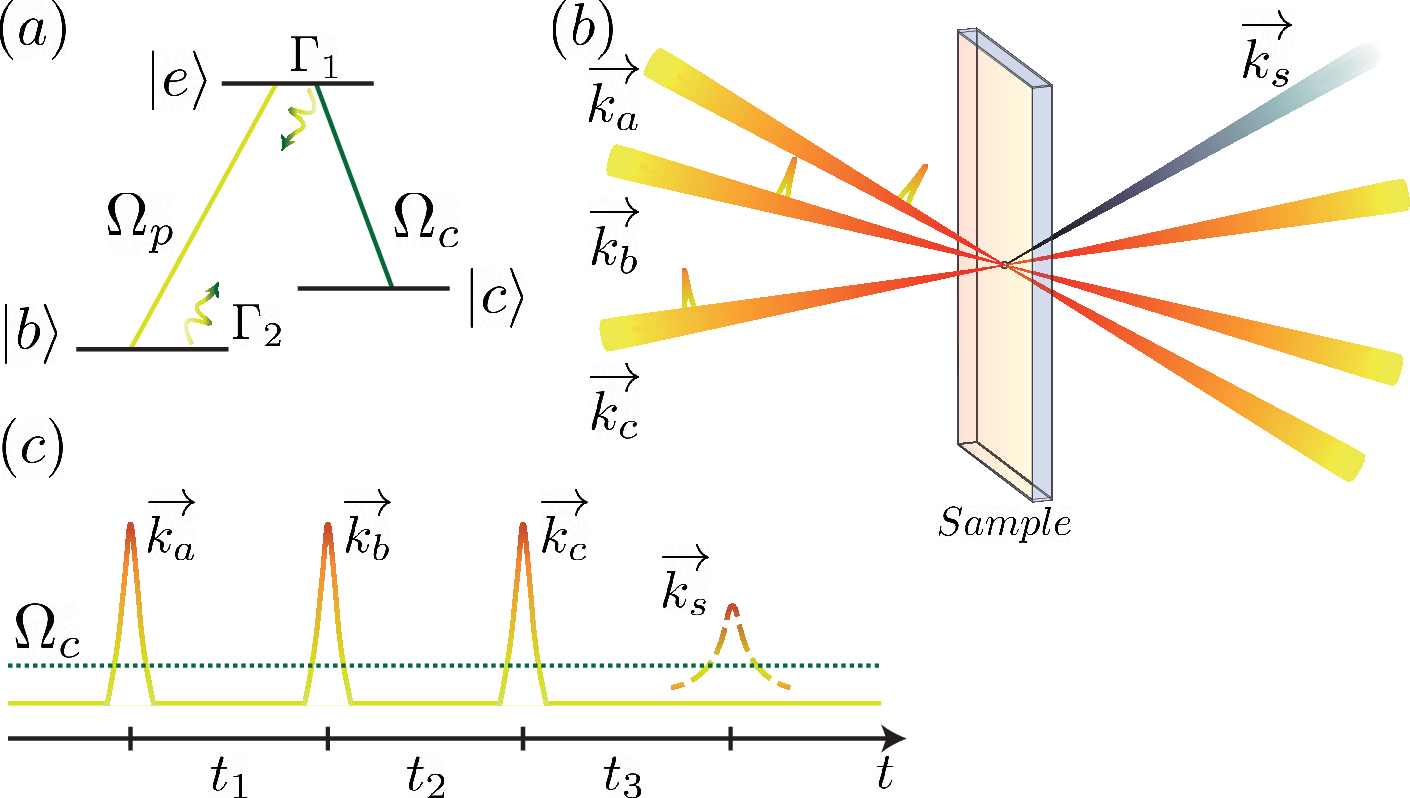}
\caption{(a) Schematic diagram of the three-level system. The energy levels $b$, $e$ and $c$ are the ground, the excited and the metastable states, respectively. The probe pulses denoted as the light-green line are applied to induce the $\vert b \rangle$ $\leftrightarrow$ $\vert e \rangle$ transition with Rabi frequency $\Omega_{be}$, and the control field denoted as the dark-green line is resonantly coupled to the $\vert e \rangle$ $\leftrightarrow$ $\vert c \rangle$ transition with Rabi frequency $\Omega_{ec}$. (b) Three probe pulses and the signal  with their different directions.
(c) Sequence of probe pulses (solid line), the control field (dotted line) and the signal (dashed line), in which $t_c~(c=1,2,3)$ is the time interval between two adjacent pulses.}
\label{fig:system_A}
\end{figure}

We consider a three-level system including a ground state $\vert b \rangle$, an excited state $\vert e \rangle$, and a metastable state $\vert c \rangle$, as shown in Fig.~\ref{fig:system_A}(a). A control field with Rabi frequency $\Omega_{ec}$ is constantly applied to the system. When a probe field is introduced, the electromagnetically-induced transparency (EIT) effect is formed to suppress the dissipation due to quantum interference \cite{fleischhauer_2005}. For 2DCS, three incident pulses of the probe field induce the $\vert b \rangle$ $\leftrightarrow$ $\vert e \rangle$ transition with Rabi frequency $\Omega_{be}$, and the signals to be measured are emitted from the system in a specific direction, as shown in Fig.~\ref{fig:system_A}(b). Between the energy levels $b$ and $e$, there exist bidirectional relaxations with a downhill rate $\Gamma_{1}$ and an uphill rate $\Gamma_{2}$. Basically, the Hamiltonian of the system without probe and control fields is
\begin{align}
H_{0}=\sum_{j}\omega_{j}\vert j\rangle\langle j\vert,
\end{align}
$(j=b,e,c)$ where we assume $\hbar=1$ for simplicity.

Considering the external fields, $H_\textrm{f}(t)=H_0+H_\textrm{int}^c(t)$ is the total Hamiltonian including only the control field, and $H_\textrm{p}(t)=H_0+H_\textrm{int}^c(t)+H_\textrm{int}^p(t)$ is the total Hamiltonian including both the control field and the probe pulses. The interaction Hamiltonian $H_{\rm{int}}^c(t)$ is written as
\begin{align}
    H_{\rm{int}}^c(t)=\Omega_{c}(t)\vert e\rangle\langle c\vert+{\rm h.c.},
    \label{Hint_RF}
\end{align}
where $\Omega_{c}(t)=-\Omega_{ec}\exp(-i\nu_{c}t)/2$, with $\Omega_{ec}$ and $\nu_{c}$ being the Rabi frequency and the frequency of the control field respectively. On the other hand, the probe pulses are strong enough, as compared to the control field and the relaxations, to have the approximation $H_\textrm{p}(t)\simeq H_0+H_\textrm{int}^p(t)~(p=a,b,c)$. Meanwhile, the probe pulses are short enough to be considered as perturbations to the system. The Hamiltonian contributed by the probe pulses will be provided later in the NHH method when it is used.

The 2DCS are plotted using the third-order RF after double Fourier transformation. This function equals the third-order polarization $P^{(3)}(t)=\langle \mu \rho^{(3)}(t) \rangle$ in the semi-impulsive limit.
Before considering the control field in $t_c$, the RF is expressed as \cite{mukamel-1995,cho-2009,hamm-2011,Tanimura1993,Tanimura2006}
\begin{eqnarray}
    &&S^{(3)}(t_{3},t_{2},t_{1})	=\left(-i\right)^{3}\times\notag \\
    &&\langle\mu^{I}\left(t_{3}+t_{2}+t_{1}\right) 	\left[\mu^{I}\left(t_{2}+t_{1}\right),\left[\mu^{I}\left(t_{1}\right),\left[\mu^{I}\left(0\right),\rho^I\left(0\right)\right]\right]\right]\rangle,\notag \\
\end{eqnarray}
where $t_{1}$, $t_{2}$ and $t_{3}$ are the time intervals between two adjacent pulses, as shown in Fig.~\ref{fig:system_A}(c). $\rho^I(0)$ represents the initial equilibrium state of the system under the Hamiltonian $H_{0}$ in the interaction picture. Meanwhile, $\mu^{I}(t)=e^{iH_{0}t}\mu e^{-iH_{0}t}$ is the electric-dipole operator in the interaction picture, where $\mu=\mu_{be}\vert b\rangle\langle e\vert + \mu_{eb}\vert e\rangle\langle b\vert$ is the electric-dipole operator in the Schr\"odinger picture. By expanding the commutators, the third-order polarization is divided into eight terms with \cite{mukamel-1995,cho-2009,hamm-2011,Tanimura2006}
\begin{equation}
\begin{split} \label{eq:Res_R}
    R_{1} =& \langle \mu^{I}(t_{3}+t_{2}+t_{1}) \mu^{I}(0) \rho^I(0) \mu^{I}(t_{1}) \mu^{I}(t_{2}+t_{1}) \rangle,  \\
    R_{2} =& \langle \mu^{I}(t_{3}+t_{2}+t_{1}) \mu^{I}(t_{1}) \rho^I(0) \mu^{I}(0) \mu^{I}(t_{2}+t_{1}) \rangle,  \\
    R_{3} =& \langle \mu^{I}(t_{3}+t_{2}+t_{1}) \mu^{I}(t_{2}+t_{1}) \rho^I(0) \mu^{I}(0) \mu^{I}(t_{1}) \rangle,  \\
    R_{4} =& \langle \mu^{I}(t_{3}+t_{2}+t_{1}) \mu^{I}(t_{2}+t_{1}) \mu^{I}(t_{1}) \mu^{I}(0) \rho^I(0) \rangle.
\end{split}
\end{equation}
and their hermitian conjugates. The incidence of three probe pulses from different directions allows for selective observation of certain combinations of the eight terms mentioned above. For example, the RP signal is emitted in the direction $\vec{k}_s=-\vec{k}_a+\vec{k}_b+\vec{k}_c$ as \cite{mukamel-1995,cho-2009,hamm-2011}
\begin{align}  \label{Eq:RPsignal_ori}
    S^{(3)}_{\rm RP}(\omega_{3},t_{2},\omega_{1})  \propto
     {\rm Re}\left[R_{2}\left(\omega_{3},t_{2},\omega_{1}\right)+R_{3}\left(\omega_{3},t_{2},\omega_{1}\right)\right].
\end{align}

There are other methods to solve the signal starting from the RF, such as the generating function \cite{Tanimura1993,Tanimura2006}, and here we present the approach using the Green function.
$R_n$ $(n=1,2,3,4)$ and their hermitian conjugates, can be obtained by solving the quantum master equation. This equation describes quantum dynamics of the open system and is written as \cite{scully-1997}
\begin{align}
    \dot{\rho}=&-i\left[H,\rho\right]-\Gamma_{1}\mathfrak{L}\left(A_{be}\right)\rho-\Gamma_{2}\mathfrak{L}\left(A_{eb}\right)\rho \notag \\
    &-\sum_{j}\gamma_{j}^{\left(0\right)}\mathfrak{L}\left(A_{jj}\right)\rho, \label{eq:ME}
\end{align}
where $H$ is the total Hamiltonian, $A_{ij}=\vert i\rangle\langle j\vert$ is the quantum jump operator describing the process of jumping from $\vert j \rangle$ to $\vert i \rangle$, $\gamma_{j}^{\left(0\right)}$ is the pure-dephasing noise for each state, and $\mathfrak{L}\left(A_{ij}\right)\rho=\frac{1}{2}\left\{ A_{ij}^{\dagger}A_{ij},\rho\right\} -A_{ij}\rho A_{ij}^{\dagger}$ with $\left\{ A_{ij}^{\dagger}A_{ij},\rho\right\}=A_{ij}^{\dagger}A_{ij}\rho+\rho A_{ij}^{\dagger}A_{ij}$ being the anti-commutator. Assuming that the control field is in resonance with the $\vert e \rangle \leftrightarrow \vert c \rangle$ transition, the time-dependent term $e^{-i\nu_{c}t}$ in Eq.~(\ref{Hint_RF}) is eliminated after the transformation to the rotating frame, i.e., the interaction picture. Thus, the time evolution of the elements of the density matrix in the interaction picture is as follows 
\begin{equation}\label{eq:MasterEq}
\begin{split}
\dot\rho^I_{bb}	=&\Gamma_{1}\rho^I_{ee}-\Gamma_{2}\rho^I_{bb},\\
\dot\rho^I_{ee}	=&\frac{i}{2}\Omega_{ec}\left(\rho^I_{ce}-\rho^I_{ec}\right)-\Gamma_{1}\rho^I_{ee}+\Gamma_{2}\rho^I_{bb},\\
\dot\rho^I_{cc}	=&-\frac{i}{2}\Omega_{ec}\left(\rho^I_{ce}-\rho^I_{ec}\right),\\
\dot\rho^I_{eb}	=&\frac{i}{2}\Omega_{ec}\rho^I_{cb}-\gamma_{eb}\rho^I_{eb},\\
\dot\rho^I_{be}	=&-\frac{i}{2}\Omega_{ec}\rho^I_{bc}-\gamma_{eb}\rho^I_{be},\\
\dot\rho^I_{ec}	=&\frac{i}{2}\Omega_{ec}\left(\rho^I_{cc}-\rho^I_{ee}\right)-\gamma_{ec}^+\rho^I_{ec},\\
\dot\rho^I_{ce}	=&-\frac{i}{2}\Omega_{ec}\left(\rho^I_{cc}-\rho^I_{ee}\right)-\gamma_{ec}^+\rho^I_{ce},\\
\dot\rho^I_{cb}	=&\frac{i}{2}\Omega_{ec}\rho^I_{eb}-\gamma_{bc}\rho^I_{cb},\\
\dot\rho^I_{bc}	=&-\frac{i}{2}\Omega_{ec}\rho^I_{be}-\gamma_{bc}\rho^I_{bc},
\end{split}
\end{equation}
where the dissipation of the energy level $c$ is neglected, and the dephasing rates read
\begin{align}
\gamma_{eb}&=\frac{1}{2}\left(\Gamma_{1}+\Gamma_{2}+\gamma_{e}^{\left(0\right)}+\gamma_{b}^{\left(0\right)}\right), \\ \gamma_{ec}^+&=\frac{1}{2}\left(\Gamma_{1}+\gamma_{e}^{\left(0\right)}+\gamma_{c}^{\left(0\right)}\right), \\ \gamma_{bc}&=\frac{1}{2}\left(\Gamma_{2}+\gamma_{b}^{\left(0\right)}+\gamma_{c}^{\left(0\right)}\right).
\end{align}

\begin{figure}[b]
\includegraphics[width=8.45cm]{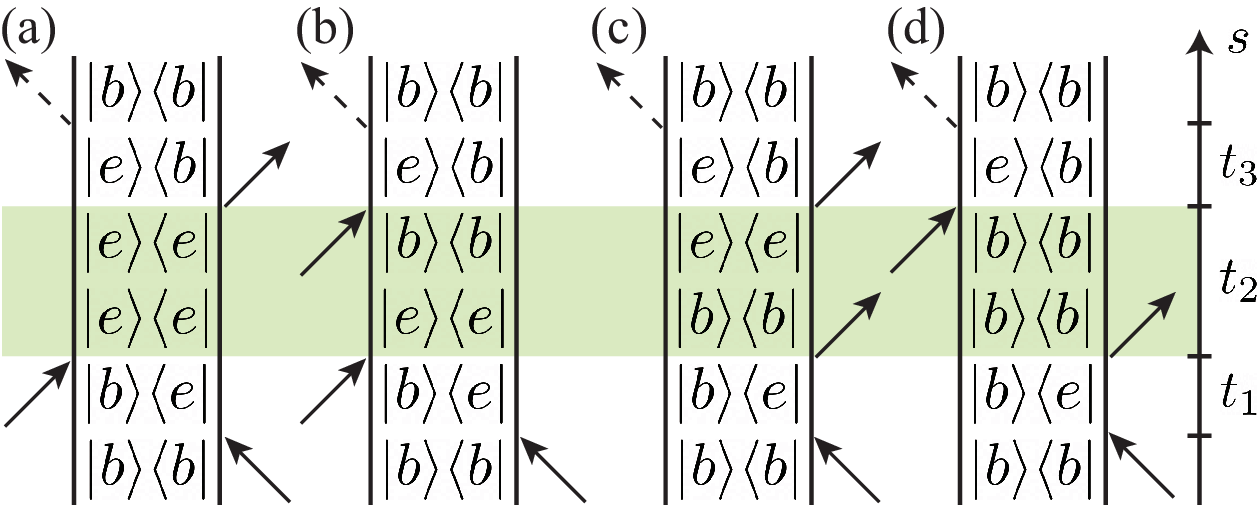}
\caption{Feynman diagrams of the three-level system by the RF formalism. (a), (b), (c) and (d) demonstrate the four Liouville paths of the RP signal, i.e., the four terms in Eq.~(\ref{Eq:RPsignal_resFun}). The evolution during $t_2$ is highlighted.}
\label{fig:FeynmanDiagram}
\end{figure}

When the non-diagonal elements $\rho_{ij}~(i\neq j)$ of the density matrix are treated perturbatively, and the Green functions are used to describe the time evolution of the density matrix elements, the RP signal is written in a more specific form \cite{liu-2020}
\begin{eqnarray} \label{Eq:RPsignal_resFun}
    S^{(3)}_{\rm RP}(\omega_{3},t_{2},\omega_{1})\!&=&\!\tilde{\mathcal{G}}_{ebeb}(\omega_{3})\mathcal{G}_{eeee}(t_{2})\tilde{\mathcal{G}}_{bebe}(\omega_{1})\notag \\
    \!&&\!+\tilde{\mathcal{G}}_{ebeb}(\omega_{3})\mathcal{G}_{bbee}(t_{2})\tilde{\mathcal{G}}_{bebe}(\omega_{1})\notag \\
    \!&&\!+\tilde{\mathcal{G}}_{ebeb}(\omega_{3})\mathcal{G}_{eebb}(t_{2})\tilde{\mathcal{G}}_{bebe}(\omega_{1})\notag \\
    \!&&\!+\tilde{\mathcal{G}}_{ebeb}(\omega_{3})\mathcal{G}_{bbbb}(t_{2})\tilde{\mathcal{G}}_{bebe}(\omega_{1}),
\end{eqnarray}
where the Green function $\mathcal{G}_{ijkl}(t)=\rho^I_{ij}(t)$ for the initial condition $\rho^I_{mn}(0)=\delta_{mk}\delta_{nl}$. The  double-sided Feynman diagrams for each term in Eq.~(\ref{Eq:RPsignal_resFun}) are shown in Fig.~\ref{fig:FeynmanDiagram}, with (a), (b), (c) and (d) corresponding to four terms in turn. The evolutions during $t_2$ with the initial and final states are explicitly demonstrated in the diagrams. The first and the last terms of Eq.~(\ref{Eq:RPsignal_resFun}) respectively correspond to $R_{2}$ and $R_{3}$ in Eq.~(\ref{eq:Res_R}). The other two terms, however, do not correspond to $R_{n}$ because they are induced by the population transfer during $t_2$. The evolution of the non-diagonal elements in $t_1$ and $t_3$ are described by the Green functions as
\begin{equation}
\begin{split}
    \tilde{\mathcal{G}}_{bebe}(\omega_{1})\!\!&=
    \!\!\frac{4(\ensuremath{\omega}_{1}-i\gamma_{bc}-\ensuremath{\omega}_{eb})}{4(\omega_{1}-i\gamma_{bc}-\omega_{eb})(\omega_{1}-i\gamma_{eb}-\ensuremath{\omega}_{eb})-\Omega_{ec}^{2}}, \\
    \tilde{\mathcal{G}}_{ebeb}(\omega_{3})\!\!&=
    \!\!\frac{4(\omega_{3}+i\gamma_{bc}-\omega_{eb})}{4(\ensuremath{\omega}_{3}+i\gamma_{bc}-\ensuremath{\omega}_{eb})(\ensuremath{\omega}_{3}+i\gamma_{eb}-\ensuremath{\omega}_{eb})-\Omega_{ec}^{2}},
\end{split}
\end{equation}
while the evolution in $t_2$ is obtained by assuming two possible initial states after the first two pulses. In the situation where the system is at $\vert e \rangle$ at the beginning of the population time, and the non-diagonal elements of the density matrix are treated as perturbative, the Green functions for the system to remain at $\vert e \rangle$ or to evolve to $\vert b \rangle$ are respectively written as
\begin{align}
    \mathcal{G}_{eeee}(t_{2})=&e^{-\frac{1}{2}\gamma_{ec}^+t_{2}}\left[\frac{A_{2}\ensuremath{\Gamma}_{2}\ensuremath{\gamma}_{ec}+\Omega_{ec}^{2}\left(\ensuremath{\gamma}_{ec}-2\Gamma_{1}\right)}{4\widetilde{\Omega}_{+}A_{1}}\sin\widetilde{\Omega}_{+}t_{2}\right. \nonumber\\    &\left.+\frac{A_{2}\Gamma_{2}+\Omega_{ec}^{2}}{2A_{1}}\cos\widetilde{\Omega}_{+}t_{2}\right]+\frac{\ensuremath{\Gamma}_{2}}{2\left(\ensuremath{\Gamma}_{1}+\ensuremath{\Gamma}_{2}\right)}  \nonumber\\
    &+\frac{\Gamma_{1}\left(2A_{1}-\Omega_{ec}^{2}\right)}{2A_{1}\left(\ensuremath{\Gamma}_{1}+\ensuremath{\Gamma}_{2}\right)}e^{-\left(\Gamma_{1}+\Gamma_{2}\right)t_{2}}, \\
    \mathcal{G}_{bbee}(t_{2})=&e^{-\frac{1}{2}\gamma_{ec}^+t_{2}}\Gamma_{1}\left[\frac{A_{2}\gamma_{ec}^++2\Omega_{ec}^{2}}{4\widetilde{\Omega}_{+}A_{1}}\sin\widetilde{\Omega}_{+}t_{2}\right. \nonumber\\
    &\left.+\frac{A_{2}}{2A_{1}}\cos\widetilde{\Omega}_{+}t_{2}\right]+\frac{\ensuremath{\Gamma}_{1}}{2\left(\ensuremath{\Gamma}_{1}+\ensuremath{\Gamma}_{2}\right)} \nonumber\\
    &-\frac{\Gamma_{1}\left(2A_{1}-\Omega_{ec}^{2}\right)}{2A_{1}\left(\ensuremath{\Gamma}_{1}+\ensuremath{\Gamma}_{2}\right)}e^{-\left(\Gamma_{1}+\ensuremath{\Gamma}_{2}\right)t_{2}},
\end{align}
where
\begin{align}
A_{1}&=\left(\Gamma_{1}+\Gamma_{2}\right)\left(\Gamma_{1}+\Gamma_{2}-\gamma_{ec}^+\right)+\Omega_{ec}^{2},\\ A_{2}&=\Gamma_{1}+\Gamma_{2}-\gamma_{ec}^+,\\ \widetilde{\Omega}_{+}&=\sqrt{\Omega_{ec}^{2}-\frac{1}{4}(\gamma_{ec}^+)^{2}}.
\end{align}
Similarly, in the other situation where the system is still at $\vert b \rangle$ after excitation, the Green functions are written as
\begin{eqnarray}
    \mathcal{G}_{eebb}(t_{2})&=&\frac{\ensuremath{\Gamma}_{2}-\ensuremath{\Gamma}_{2}e^{-\left(\ensuremath{\Gamma}_{1}+\ensuremath{\Gamma}_{2}\right)t_{2}}}{\Gamma_{1}+\Gamma_{2}}, \\
    \mathcal{G}_{bbbb}(t_{2})&=&\frac{\Gamma_{1}+\Gamma_{2}e^{-\left(\Gamma_{1}+\Gamma_{2}\right)t_{2}}}{\Gamma_{1}+\Gamma_{2}}.\label{eq:Gbbbb}
\end{eqnarray}
Eventually, by taking the real part of the double Fourier transformed RF, the RP signal and thus the 2DCS are obtained.

\subsection{non-Hermitian Hamiltonian method} \label{sec:Non-Hermitian Method}

We apply the NHH method to the same system in Fig.~\ref{fig:system_A}(a).  Starting directly from the master Eq.~(\ref{eq:ME}), we drop the quantum-jump terms and approximately obtain
\begin{align} \label{eq:master equation for non-H}
    \dot{\rho}=-i(H_{\rm NH}\rho-\rho H_{\rm NH}^\dagger ),
\end{align}
where the NHH reads \cite{Ai2021}
\begin{align}
    H_{\rm NH}=H-\frac{i}{2}(\Gamma_{1}A_{be}^{\dagger}A_{be}+\Gamma_{2}A_{eb}^{\dagger}A_{eb}+\sum_{j}\gamma_{j}^{(0)}A_{jj}^{\dagger}A_{jj}).
\end{align}
The NHH method is valid if two conditions are met. The first is that the dissipation and dephasing rates $\Gamma_{j}$ and $\gamma_{j}^{(0)}$ should be much smaller than the typical frequency of the Hamiltonian, while the other would be discussed later.

By the NHH method, the final state of the system in the interaction picture is written as \cite{Chen-2017,Cai2022PRL}
\begin{align}
    &\vert\psi^I(t_{3},t_{2},t_{1})\rangle=U^{I}_{f}(t_{1}+t_{2}+t_{3}+\delta t_{a}+\delta t_{b}+\delta t_{c}, \notag \\
    &t_{1}+t_{2}+\delta t_{a}+\delta t_{b}+\delta t_{c})\notag\\
    &\times U^{I}_p\left(t_{1}+t_{2}+\delta t_{a}+\delta t_{b}+\delta t_{c},t_{1}+t_{2}+\delta t_{a}+\delta t_{b}\right)\notag\\
    &\times U^{I}_{f}\left(t_{1}+t_{2}+\delta t_{a}+\delta t_{b},t_{1}+\delta t_{a}+\delta t_{b}\right)\notag\\
    &\times U^{I}_p\left(t_{1}+\delta t_{a}+\delta t_{b},\tau+\delta t_{a}\right)U^{I}_{f}\left(t_{1}+\delta t_{a},\delta t_{a}\right)\notag\\
    &\times U^{I}_p\left(\delta t_{a},0\right)\vert\psi^I\left(0\right)\rangle,
\end{align}
where $\delta t_p$ ($p=a,b,c$) is the time interval of the probe pulse, and $U^{I}_p$ ($U^{I}_f$) represents the evolution operator with (without) the probe pulses included in the interaction picture. As in the RF formalism, we consider that the probe pulses are strong enough relative to the control field, and their durations are short enough that both dissipation and dephasing during the excitations are neglected. As a result, the Hamiltonians during the probe pulses are explicitly written in the interaction picture as
\begin{align}
    H^I_{p}=&-\frac{1}{2}\Omega_{be}e^{-i\vec{k}_{p}\cdot\vec{r}}\vert b\rangle\langle e\vert+{\rm h.c.}
\end{align}
 Taking dissipation and dephasing into account, the Hamiltonians during the free evolutions read
\begin{align}
    H^I_{f}=&-\frac{1}{2}\Omega_{ec}(\vert e\rangle\langle c\vert+\vert c\rangle\langle e\vert)\notag\\
    &-\frac{i}{2}(\Gamma_{1}\vert e\rangle\langle e\vert+\Gamma_{2}\vert b\rangle\langle b\vert+\sum_{j}\gamma_{j}^{(0)}\vert j\rangle\langle j\vert).
\end{align}
The Rabi frequency $\Omega_{ij}$ is given by $-\vec{d}_{ij}\cdot\vec{E}$, where $\vec{E}$ is the electric field of the laser and $\vec{d}_{ij}$ is the transition dipole moment between state $i$ and $j$. Thus, the selectivity through laser polarization is accounted for and directly reflected in $\Omega_{ij}$.

Hereinto, we approximate the probe pulses as square pulses. Under $H^I_f$, the time evolutions $U_{f}^{I}(t_c+t_0,t_0)=e^{iH_{0}(t_c+t_0)}U_{f}^{S}(t_c+t_0,t_0) e^{-iH_{0}t_0}=e^{iH_{0}t_c}U_{f_I}(t_c)$, where $t_0$ represents the starting time of the free evolution, and $U_{f}^{S}$ is the time evolution operator in the Schr\"{o}dinger picture. Then $U_{f_I}(t_c)$ for initial states $\vert b \rangle$, $\vert e \rangle$ and $\vert c \rangle$ are respectively written as
\begin{align}
    U_{f_I}(t_c)\vert b\rangle&=C_{bb}(t_c)\vert b\rangle, \\
    U_{f_I}(t_c)\vert e\rangle&=C_{ee}(t_c)\vert e\rangle+C_{ce}(t_c)\vert c\rangle, \\
    U_{f_I}(t_c)\vert c\rangle&=C_{ec}(t_c)\vert e\rangle+C_{cc}(t_c)\vert c\rangle,
\end{align}
 and
\begin{align}
    C_{bb}(t)&=e^{-\frac{1}{2}\gamma_{b}t}, \label{Eq:Uf_b}\\
    C_{ee}(t)&=e^{-\frac{1}{2}\gamma_{ec}^+t}\frac{(2i\tilde{\Omega}_{-}-\gamma_{ec}^-)e^{\frac{i}{2}\tilde{\Omega}_{-}t}+(2i\tilde{\Omega}_{-}+\gamma_{ec}^-)e^{-\frac{i}{2}\tilde{\Omega}_{-}t}}{4i\tilde{\Omega}_{-}} \\
    C_{ce}(t)&=e^{-\frac{1}{2}\gamma_{ec}^+t}\frac{\Omega_{ec}\left(e^{i\tilde{\Omega}_{-}t}-1\right)e^{-\frac{i}{2}\tilde{\Omega}_{-}t}}{2\tilde{\Omega}_{-}}, \\
    C_{ec}(t)&=e^{-\frac{1}{2}\gamma_{ec}^+t}\frac{\Omega_{ec}\left(e^{i\tilde{\Omega}_{-}t}-1\right)e^{-\frac{i}{2}\tilde{\Omega}_{-}t}}{2\tilde{\Omega}_{-}}, \\
    C_{cc}(t)&=e^{-\frac{1}{2}\gamma_{ec}^+t}\frac{(2i\tilde{\Omega}_{-}-\gamma_{ec}^-)e^{-\frac{i}{2}\tilde{\Omega}_{-}t}+(2i\tilde{\Omega}_{-}+\gamma_{ec}^-)e^{\frac{i}{2}\tilde{\Omega}_{-}t}}{4i\tilde{\Omega}_{-}}. \label{Eq:Uf_c}
\end{align}
In the subscript of $C_{ji}(t)$, $i$ and $j$ represent the initial and final state during the evolution, and
\begin{align}
    \gamma_{b}=&\Gamma_{2}+\gamma_{b}^{(0)}, \\
    \gamma_{ec}^-=&\Gamma_{1}+\gamma_{e}^{(0)}-\gamma_{c}^{(0)},\\
    \tilde{\Omega}_{-}=&\sqrt{\Omega_{ec}^{2}-\frac{1}{4}(\gamma_{ec}^-)^{2}}.
\end{align}
 Meanwhile, considering $\delta t_{p}\simeq 0$, the time evolutions $U_{p}^{I}(\delta t_{p}+t_0,t_0)=U_{p_I}(\delta t_{p})$ with $t_0$ denoting the starting time of the probe pulses. The $U_{p_I}(\delta t_{p})$ under $H^I_p$ for three initial states are then respectively written as
\begin{align}
    U_{p_I}(\delta t_{p})\vert b\rangle=&N_{p}\left(\vert b\rangle+\beta_{p}e^{i\vec{k_{p}}\vec{r}}\vert e\rangle\right), \label{Eq:Up_b}\\
    U_{p_I}(\delta t_{p})\vert e\rangle=&N_{p}\left(\beta_{p}e^{-i\vec{k_{p}}\vec{r}}\vert b\rangle+\vert e\rangle\right), \\
    U_{p_I}(\delta t_{p})\vert c\rangle=&\vert c\rangle,
\end{align}
where $\beta_{p}=i\Omega_{be}\delta t_{p}/2$, and the normalization constants are
\begin{align*}
    N_{p}=\left(1+\left|\beta_{p}\right|^{2}\right)^{-1/2}.
\end{align*}

Straightforward, the final density matrix reads $\rho^I(t)=\vert\psi^I\left(t_{3},t_{2},t_{1}\right)\rangle \langle\psi^I\left(t_{3},t_{2},t_{1}\right)\vert$, and the total polarization is $P(t)=\langle \mu^I(t) \rho^I(t) \rangle$. Under the situation that the spacing of the energy levels are much wider than their linewidths, its non-diagonal elements $\rho^I_{eb}$ and $\rho^I_{be}$ emit signals near $\omega_e$ in the frequency domain, which are proportional to the polarization 
\begin{align}
    \vec{P}_{eb}(t_{3},t_{2},t_{1})=\rho_{eb}^{I}(t_{3},t_{2},t_{1}){\mu}^{I}_{be}(t_{1}+t_{2}+t_{3}+\delta t_{a}+\delta t_{b}+\delta t_{c})+\rm c.c.
\end{align}
Here, ${\mu}^{I}_{be}(t)$ is the component corresponding to the transition from $\vert e \rangle$ to $\vert b \rangle$ of the electric-dipole operator ${\mu}^I(t)$. On account of the phase-matching condition, the polarization $\vec{P}_{\rm RP}(t_{3},t_{2},t_{1})$ in the RP direction along $\vec{k}_{s}=-\vec{k}_{a}+\vec{k}_{b}+\vec{k}_{c}$ is selected. Furthermore, by double Fourier transforms with respect to $t_1$ and $t_3$, the RP signal of the 2DCS reads
\begin{align}
    \vec{P}_{\rm RP}(\omega_{3},t_{2},\omega_{1})=\mathcal{F}\left[\vec{P}_{\rm RP}(t_{3},t_{2},t_{1})\right],
\end{align}
and is explicitly written as
\begin{align} \label{Eq:RPsignal_NH}
    \vec{P}_{\rm RP}(\omega_{3},t_{2},&\omega_{1})= N_{a}^{2}N_{b}N_{c}\beta_{a}^{*}\times\notag \\
    &[N_{b}N_{c}\beta_{b}^{*}\beta_{c}
    \tilde{G}_{bebe}(\omega_{1})G_{bbbb}(t_{2})\tilde{G}_{ebeb}(\omega_{3}) \notag \\
    &+N_{b}N_{c}\beta_{b}\beta_{c}^{*}\tilde{G}_{bebe}(\omega_{1})G_{eeee}(t_{2})\tilde{G}_{ebeb}(\omega_{3}) \notag \\
    &+N_{b}\beta_{b}\beta_{c}^{*}\tilde{G}_{bebe}(\omega_{1})G_{ceee}(t_{2})\tilde{G}_{ebcb}(\omega_{3}) \notag \\
    &+N_{c}\beta_{b}\beta_{c}^{*}\tilde{G}_{bcbe}(\omega_{1})G_{eeec}(t_{2})\tilde{G}_{ebeb}(\omega_{3}) \notag \\
    &+\beta_{b}\beta_{c}^{*}\tilde{G}_{bcbe}(\omega_{1})G_{ceec}(t_{2})\tilde{G}_{ebcb}(\omega_{3})].
\end{align}
Here, $\exp(-i\omega_{e}t_{c})$ are absorbed into $G_{ijkl}(t)$ and $\tilde{G}_{ijkl}(\omega)$. The quasi-Green functions and their Fourier transform $G_{ijkl}(t)$ and $\tilde{G}_{ijkl}(\omega)$ are proposed for the NHH method, and are written as
\begin{align}
G_{ijkl}(t)=&C_{jl}^{*}(t)C_{ik}(t)e^{-i\omega_{e}t},\\
\tilde{G}_{ijkl}(\omega)=&\mathcal{F}\left[G_{ijkl}(t)\right], \end{align}
where $kl$ represent the initial density matrix element $\vert k \rangle\langle l \vert$ before an evolution, and $ij$ represent the final density matrix element $\vert i \rangle\langle j \vert$ after the evolution. Since $C_{ji}$ ($C_{ji}^{*}$) are components of the ket (bra), we rearrange the subscripts of $C_{jl}^{*}C_{ik}$ as $G_{ijkl}$ analogue to those of the Green functions. After the Fourier transform, the quasi-Green functions read
\begin{align}
\tilde{G}_{bebe}(\omega)&=\frac{-\gamma_{ec}^-+2\left[\gamma_{b}+\gamma_{ec}^++2i\left(\omega-\omega_{e}\right)\right]}{\tilde{\Omega}_{-}^{2}+\left[\gamma_{b}+\gamma_{ec}^++2i\left(\omega-\omega_{e}\right)\right]^{2}}, \\
\tilde{G}_{bcbe}(\omega)&=\frac{2i\Omega_{ec}}{\tilde{\Omega}_{-}^{2}+\left[\gamma_{b}+\gamma_{ec}^++2i\left(\omega-\omega_{e}\right)\right]^{2}}, \\
\tilde{G}_{ebeb}(\omega)&=\frac{-\gamma_{ec}^-+2\left[\gamma_{b}+\gamma_{ec}^+-2i\left(\omega-\omega_{e}\right)\right]}{\tilde{\Omega}_{-}^{2}+\left[\gamma_{b}+\gamma_{ec}^+-2i\left(\omega-\omega_{e}\right)\right]^{2}}, \\
\tilde{G}_{ebcb}(\omega)&=\frac{2i\Omega_{ec}}{\tilde{\Omega}_{-}^{2}+\left[\gamma_{b}+\gamma_{ec}^+-2i\left(\omega-\omega_{e}\right)\right]^{2}}.
\end{align}

\begin{figure}[h]
\includegraphics[width=8.45cm]{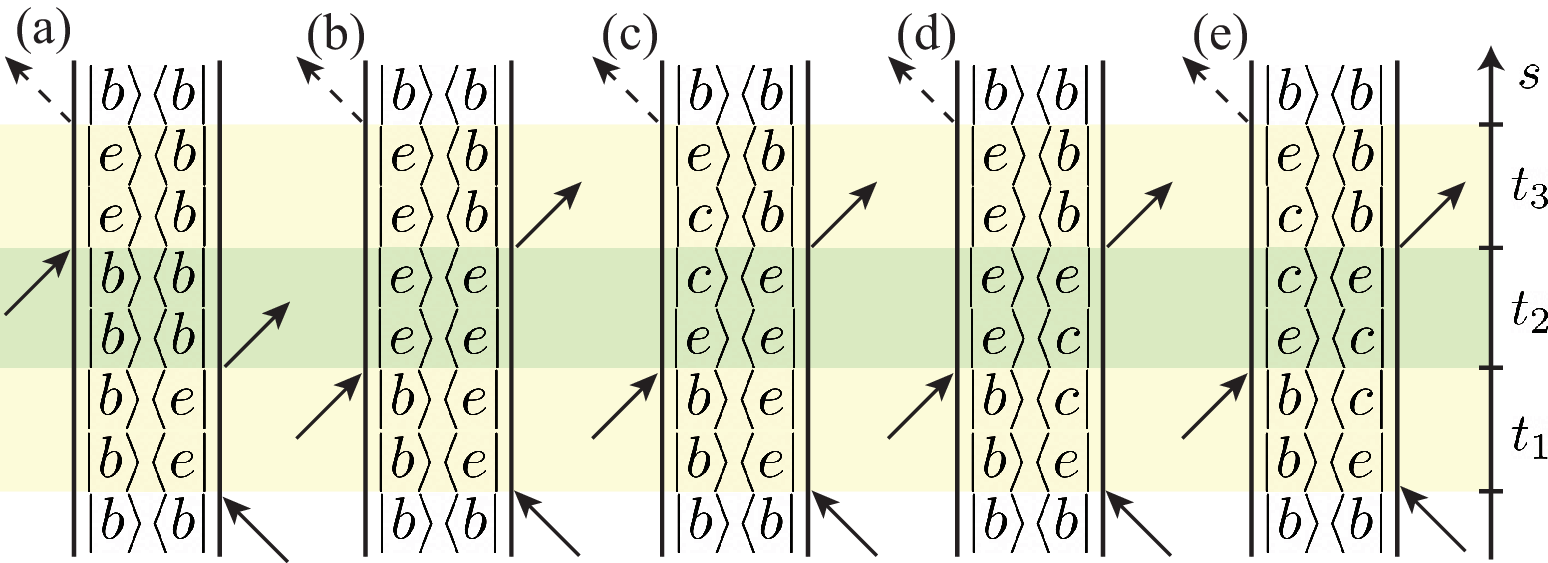}
\caption{Feynman diagrams of the three-level system by the NHH method. (a)-(e) demonstrate the five Liouville paths of the RP signal, i.e., the five terms in Eq.~(\ref{Eq:RPsignal_NH}). The evolution during $t_2$ is highlighted by green, and the evolutions during $t_1$ and $t_3$ are marked by yellow.}
\label{fig:FeynmanDiagram_NH}
\end{figure}

The Feynman diagrams of each term in Eq.~(\ref{Eq:RPsignal_NH}) are shown in Fig.~\ref{fig:FeynmanDiagram_NH}, with (a)-(e) corresponding to the five terms in turn. The initial and final states of evolutions during $t_1$, $t_2$ and $t_3$ are explicitly given. Compared with Fig.~\ref{fig:FeynmanDiagram}, two of the five Liouville paths are the same, i.e., Fig.~\ref{fig:FeynmanDiagram_NH}(a) vs Fig.~\ref{fig:FeynmanDiagram}(d) and Fig.~\ref{fig:FeynmanDiagram_NH}(b) vs Fig.~\ref{fig:FeynmanDiagram}(a). Meanwhile, the other two Liouville paths disappear, i.e., Fig.~\ref{fig:FeynmanDiagram}(b) and (c), and three new Liouville paths emerge, i.e., Fig.~\ref{fig:FeynmanDiagram_NH}(c)-(e). More specifically, Fig.~\ref{fig:FeynmanDiagram_NH}(c) remains the same as Fig.~\ref{fig:FeynmanDiagram}(a) until the initial state of the $t_2$ evolution, and diverges at the final state of it. In addition, Fig.~\ref{fig:FeynmanDiagram_NH}(d) and (e) are different from Fig.~\ref{fig:FeynmanDiagram} at the final state of the $t_1$ evolution. We note that it is these distinctions that lead to the difference between the two methods. 

Finally, a summary of the approximations used in the RF formalism and the NHH method is presented in Table \ref{tab:approximation}, highlighting both the similarities and differences in their treatment. First of all, since the quantum-jump terms are neglected, the detailed balance can not be reproduced in the NHH method.
Additionally, to simplify the calculations, the square-pulse approximation is adopted. However, more realistic pulse shapes can be considered if $H^{p}_\textrm{int}(t)$ is appropriately revised. More importantly, in the RF formalism, we treat the off-diagonal terms of the density matrix perturbatively to obtain the analytic solution. The more-accurate results may be achieved if this approximation is discarded. 
\begin{table}[h]
{\footnotesize
\centering
\renewcommand{\arraystretch}{1}
\begin{tabular}{C{2.01cm} | C{3.22cm} C{3.22cm}}
\hline
\backslashbox{Appr.}{\hspace{-5mm} Method} & RF formalism & NHH method \\
\hline
\multirow{3}{*} {Different} & perturbative $H_\textrm{int}^p$ & square pulse appr. \\
 & semi-impulsive limit & non-Hermitian $H$ \\
 & perturbative $\rho_{ij}$ &  \\
 \hline
 \multirow{2}{*} {Same} 
 & \multicolumn{2}{c}{$H_\textrm{p}\simeq H_0+H_\textrm{int}^p$} \\
 & \multicolumn{2}{c}{$\nu_{c}=\omega_{ec}$} \\

\hline
\end{tabular}
\caption{Summary of approximations used in the two methods. ``Appr." is short for ``Approximation".}
\label{tab:approximation}
}
\end{table}

\section{Results and Discussion} \label{sec:Results and Discussion}

In this section, we compare the 2DCS of the two methods by the example of 1,2-propanediol gas at room temperature. Three states under investigation are selected with $\vert b\rangle=\vert v_g\rangle\vert 0_{0,0,0} \rangle$, $\vert e\rangle=\vert v_e\rangle\vert 1_{1,1,1} \rangle$ and $\vert c\rangle=\vert v_e\rangle\vert 1_{0,1,0} \rangle$. $\vert v_g\rangle$ ($\vert v_e\rangle$) marks the vibrational ground (first-excited) state, and $\vert J_{K_a,K_c,M}\rangle$ marks the rotational states. The transition frequencies of each states are respectively $\omega_b=0$, $\omega_e/2\pi=4.33 ~\rm THz$, and $\omega_c/2\pi=4.32 ~\rm THz$ \cite{chen-2020,arenas-2017,lovas-2009,Cai2022PRL}. The Rabi frequencies and durations of the probe pulses are $\Omega_{be}/2\pi=50~\rm MHz$ and $\delta t_p=0.5 ~\rm ns$, with a bandwidth of $2\pi \times 1.8 ~\rm GHz < \omega_{ec}$, while the Rabi frequency of the control field  is $\Omega_{ec}/2\pi=2~\rm MHz$ \cite{Cai2022PRL}. The dissipation rates are $\Gamma_1/2\pi=1~\rm kHz$ and $\Gamma_2/2\pi=0.03~\rm kHz$, and the pure-dephasing rates of all states are homogeneously $\gamma_{b}^{\left(0\right)}/2\pi=\gamma_{e}^{\left(0\right)}/2\pi=\gamma_{c}^{\left(0\right)}/2\pi=0.1~\rm MHz$ \cite{patterson-2012,eibenberger-2017}.

\begin{figure}
\includegraphics[width=8.45 cm]{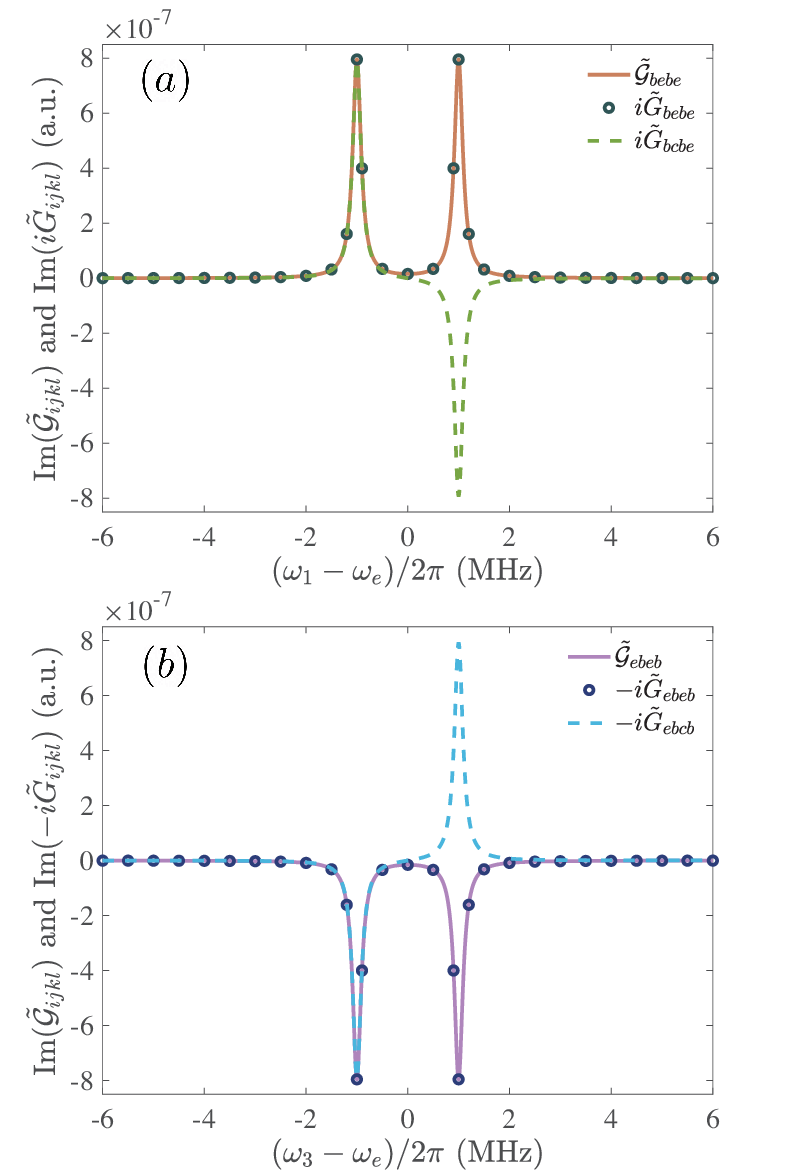}
\caption{Comparison of the Green functions by the RF ($\mathcal{\tilde{G}}_{ijkl}$) and the NHH ($\tilde{G}_{ijkl}$) approaches. (a) $\mathcal{\tilde{G}}_{bebe}$ vs $i\tilde{G}_{bebe}$, and $i\tilde{G}_{bcbe}$ are respectively denoted by the orange solid, dark-green circles, light-green dashed lines, (b) $\mathcal{\tilde{G}}_{ebeb}$ vs $-i\tilde{G}_{ebeb}$, and $-i\tilde{G}_{ebcb}$ are respectively denoted by the purple solid, dark-blue circles, light-blue dashed lines.}
\label{fig:GreenFunction_w1w3}
\end{figure}

The comparison between the Green function $\mathcal{\tilde{G}}_{ijkl}(\omega)$ and the quasi-Green functions $\tilde{G}_{ijkl}(\omega)$ are shown in Fig.~\ref{fig:GreenFunction_w1w3}. The additional factors $\pm i$ before $\tilde{G}_{ijkl}(\omega)$ are included due to the different definitions of the Fourier transforms in the RF and the NHH approaches \cite{mukamel-1995}. The curves of Im($\mathcal{\tilde{G}}_{bebe}$) and Im($i\tilde{G}_{bebe}$) are identical, and so are Im($\mathcal{\tilde{G}}_{ebeb}$) and Im($-i\tilde{G}_{ebeb}$). Concurrently, the curve of Im($i\tilde{G}_{bcbe}$) coincides with Im($\mathcal{\tilde{G}}_{bebe}$) and Im($i\tilde{G}_{bebe}$) when $\omega-\omega_e<0$, and differs from them by a sign when $\omega-\omega_e>0$. The same behavior can be observed in the curves of Im($-i\tilde{G}_{ebcb}$) vs Im($\mathcal{\tilde{G}}_{ebeb}$) and Im($-i\tilde{G}_{ebeb}$). The imaginary parts of the Green functions for $\omega_1$ dominate the absorption of probe pulses, with higher positive values representing greater absorption and lower negative values representing more emission. The opposite sign of the imaginary parts of the Green functions for $\omega_3$ have the same influence. The control field causes the original absorption peak at $\omega_{e}$ to be split into two peaks with approximately-symmetrical locations around $\omega_{e}$, just as EIT happens. Furthermore, the emergent evolution paths of $\tilde{G}_{bcbe}$ and $\tilde{G}_{ebcb}$ introduce emission peaks. Consequently, the splitting determines the location of peaks on 2DCS, and the influence of the emission peaks will be discussed later.

 \begin{figure*}[t]
 \includegraphics[width=16.5 cm]{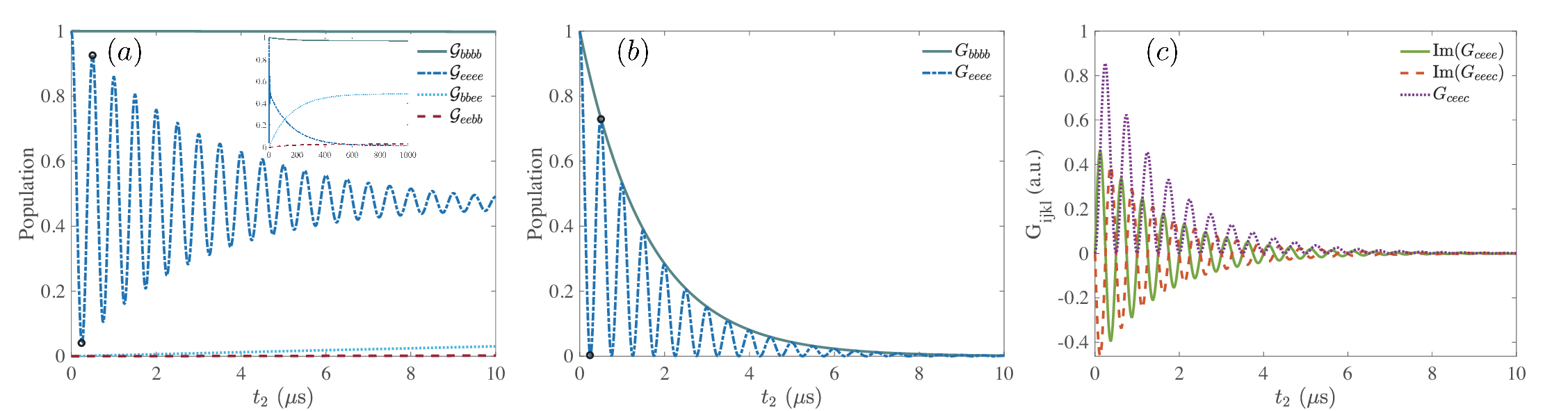}
 \caption{Population dynamics by (a) the RF formalism, (b) the NHH method and (c) the quasi-Green function $G_{ijkl}(t)$ for the coherences vs time. The dark-green solid and dark-blue dash-dotted lines denote $\mathcal{G}_{bbbb}$ ($G_{bbbb}$) and $\mathcal{G}_{eeee}$ ($G_{eeee}$), respectively. The light-blue dotted and dark-red dashed lines respectively denote $\mathcal{G}_{bbee}$ and $\mathcal{G}_{eebb}$, while the light-green solid, orange dashed, and purple dotted lines respectively denote $\textrm{Im}(G_{ceee})$, $\textrm{Im}(G_{eeec})$ and $G_{ceec}$. Here, $G_{ceee}$ and $G_{eeec}$ are purely imaginary. The inset in (a) shows the long-time regime, and the black circles in the first two subfigures label $t_2=0.24~\mu$s and $t_2=0.50~\mu$s, respectively.}
 \label{fig:Population}
 \end{figure*}

The population dynamics vs the population time $t_2$ by the RF formalism and the NHH method are compared in Fig.~\ref{fig:Population}, with only possible Liouville paths demonstrated. Figure~\ref{fig:Population}(a) illustrates the population dynamics by the RF formalism in the subspace spanned by $|b\rangle$ and $|e\rangle$. Figure~\ref{fig:Population}(b) depicts the dynamics of the diagonal elements $\vert b\rangle\langle b \vert$ and $\vert e\rangle\langle e \vert$ of the density matrix by the NHH method. Figure ~\ref{fig:Population}(c) shows the quasi-Green functions for the three Liouville paths by the NHH method in addition to the RF formalism. Notice that although the initial (final) state of $G_{ceee}$ ($G_{eeec}$) is the population state, its final (initial) state is a coherent state. This sharp contrasts yield purely-imaginary quasi-Green functions.

 The dynamic processes of $\mathcal{G}_{bbbb}$ and $G_{bbbb}$ exhibit a damping with different timescales for reaching the steady states, i.e., $(\Gamma_1+\Gamma_2)^{-1}$ vs $(\gamma_b/2)^{-1}$. For $\mathcal{G}_{eeee}$,  $G_{eeee}$ and $G_{ceec}$, there are coherent oscillations due to the coherent population transfer. The oscillation frequency of  $\mathcal{G}_{eeee}$ is $\widetilde{\Omega}_{+}$  when $\Omega_{ec}>\gamma_{ec}^+/2$. Meanwhile, the oscillation frequencies of $G_{eeee}$ and $G_{ceec}$ are identical, i.e., $\widetilde{\Omega}_{-}$ when $\Omega_{ec}>\gamma_{ec}^-/2$. These three curves all oscillate at the same frequency, since $\widetilde{\Omega}_{+} \approx \widetilde{\Omega}_{-}\approx\Omega_{ec}$ when $\Omega_{ec}\gg \gamma_{ec}^{\pm}$. On the other hand, oscillations undergo damping with a time scale of $1/\gamma_{ec}^+$, and ultimately disappear due to dephasing and dissipation.

 There is a phase shift of $\pi$ between the oscillations of $G_{eeee}$ and $G_{ceec}$. When $G_{eeee}=0$, $G_{ceec}$ reaches its maximum values, indicating that the maximum population transfer corresponds to the maximum coherence. In addition, the imaginary parts of $G_{ceee}$ and $G_{eeec}$ exhibit anti-synchronized damping oscillations. They oscillate around zero, with a phase shift of $\pi$ between them as well. However, they have no direct effect on the final 2DCS.
  Moreover, although the oscillation behaviors by the two methods exhibit similarity, they tend to reach different steady states in long-time limit, e.g., $\Gamma_1/(\Gamma_1+\Gamma_2)$ for $\mathcal{G}_{bbbb}$ vs zero for $G_{bbbb}$, according to Eqs.~(\ref{eq:Gbbbb}) and (\ref{Eq:Uf_b}). This remarkable discrepancy is due to the fact that the Hamiltonian in the NHH method is not Hermitian, i.e., $H^\dagger\neq H$, so the conservation of the trace is violated and thus the detailed balance breaks down.

\begin{figure}
\includegraphics[width=8.45 cm]{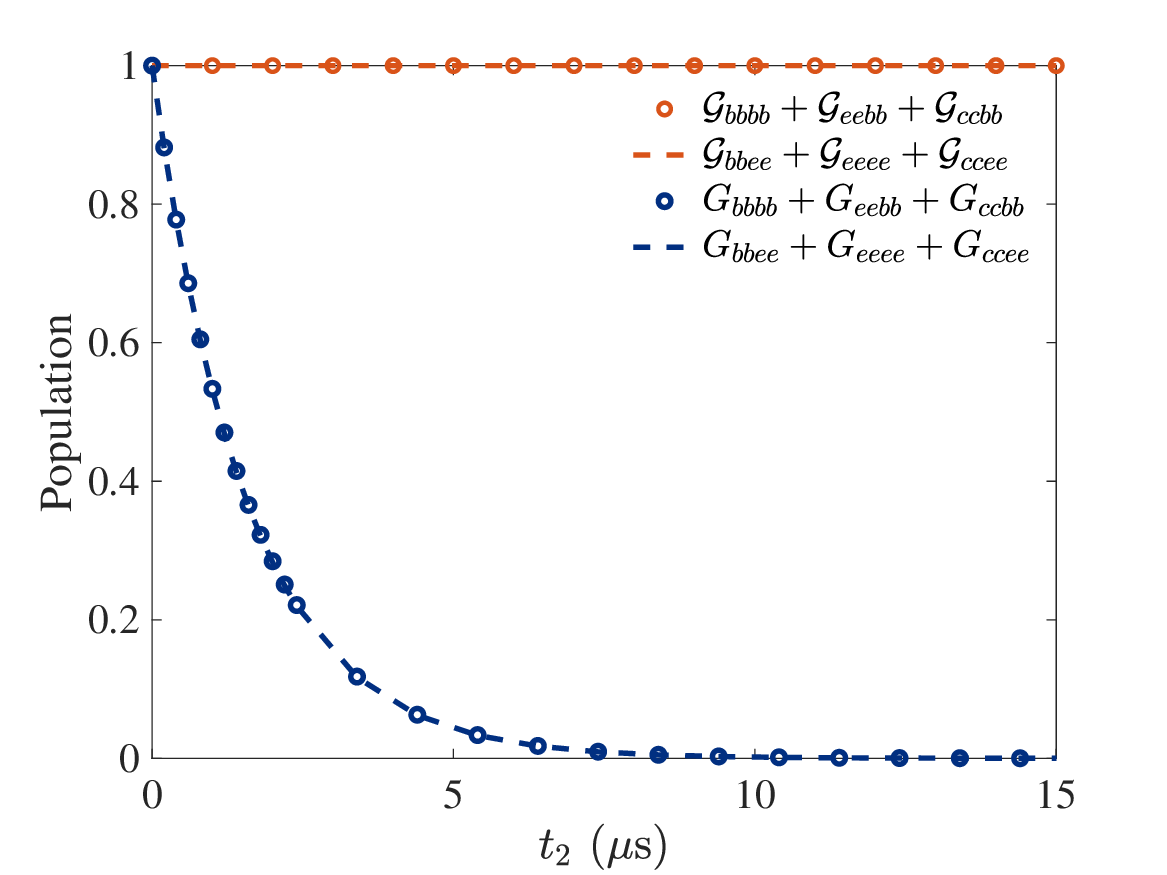}
\caption{Conservation of trace for different initial states by the RF and the NHH approaches. The orange circles and dashed lines denote the traces of the RF formalism starting from the initial states $\vert b \rangle$ and $\vert e \rangle$, and the dark-blue circles and dashed lines denote the traces of the NHH method starting from the initial states $\vert b \rangle$ and $\vert e \rangle$, respectively.}
\label{fig:PopulationSum}
\end{figure}

To illustrate this more clearly, we present a comparison of the trace dynamics during $t_2$ for the two methods in Fig.~\ref{fig:PopulationSum}, where the orange circles and dashed lines always remain at unity, while the dark-blue circles and dashed lines decay exponentially. The circles (dashed lines) represent the summation of all populations evolving initially from the energy level $b$ ($e$). The result indicates that the trace is conserved for the RF formalism no matter how long the system evolves. In contrast, the trace is not conserved for the NHH method whatever the system in $t_2$ is initially prepared.

\begin{figure*}[t]
\includegraphics[width=16.5 cm]{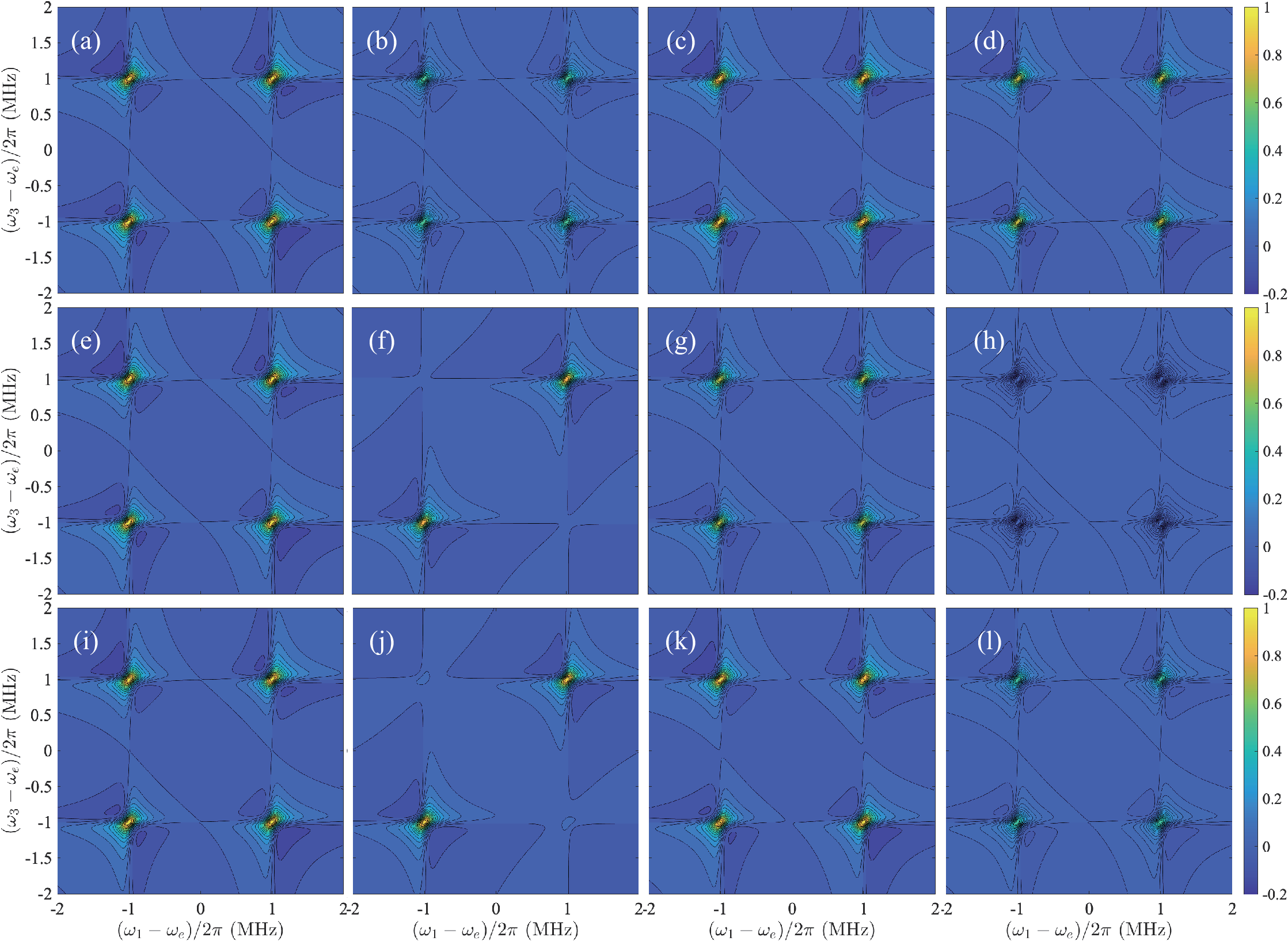}
\caption{Comparison of the 2DCS for the two methods. The first (second) row correspond to the RF formalism (the NHH method).
The third row is by the RF formalism with Liouville paths of the same as the NHH method in Fig.~\ref{fig:FeynmanDiagram_NH}. The 2DCS are plotted when (a,e,i) $t_2=0$~$\rm \mu s$, (b,f,j) $t_2=0.24$~$\rm \mu s$, (c,g,k) $t_2=0.50$~$\rm \mu s$, (d,h,l) $t_2=10^3$~$\rm \mu s$.}
\label{fig:2DPlot}
\end{figure*}

For further comparison of the two approaches, representative points are selected according to their population dynamics, i.e., $t_2=0,~0.24,~0.5,~10^3$~$\rm \mu s$, and the 2DCS for each method at these population times are plotted in Fig.~\ref{fig:2DPlot} (a) through (h). The first three $t_2$ values correspond to the initial state, the first valley and the first peak of the population dynamics, respectively, for both the RF formalism and the NHH method. The last $t_2$ is selected for observing results at the steady state. In both methods, the introduction of the control field results in splitting of the homogeneously-broadened peak into four smaller peaks on 2DCS \cite{liu-2020}. The RF formalism and the NHH method, exhibiting identical peak positions, indicate that the spacings between peaks are identical. These spacings between the adjacent diagonal and non-diagonal peaks equal to $2.005\times 2 \pi ~\rm MHz$, which are approximately the splitting width $2\times 2 \pi ~\rm MHz$ of the Green functions according to Fig.~\ref{fig:GreenFunction_w1w3}.

\begin{figure*}[h]
\includegraphics[width=16.5 cm]{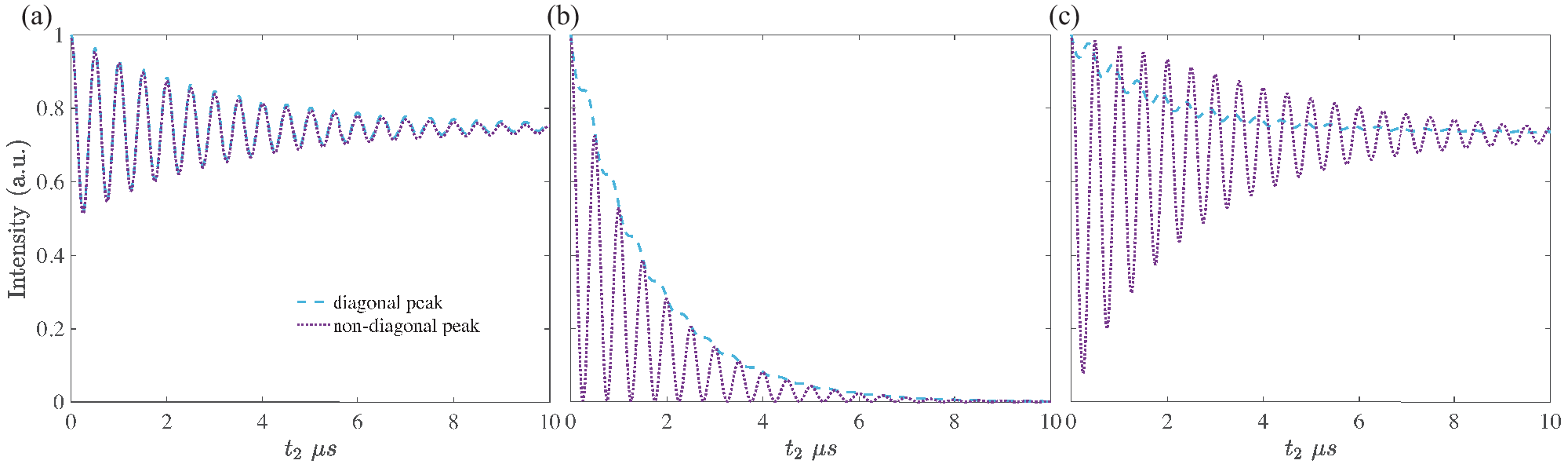}
\caption{The heights of diagonal and non-diagonal peaks are shown by (a) the RF formalism, (b) the NHH method, and (c) the RF formalism with the Liouville paths from the NHH method, as a function of $t_2$. The light-blue dashed and purple dotted lines represent the diagonal and non-diagonal peak heights, respectively.}
\label{fig:slice}
\end{figure*}

The RP signals of the two methods are respectively normalized by their maximum values at $t_2=0~\rm \mu s$, thereby obtaining relative signals for the sake of simplicity. At $t_2=0~\rm \mu s$, both methods yield the highest four peaks, and the relative heights for one peak and its counterpart from the other method are approximately identical, with a variance at the scale of $10^{-4}$. At $t_2=10^{3}~\rm \mu s$, the peaks of the RF formalism retain their scale because the trace is conserved. The peaks of the NHH method, however, undergo overall attenuation as $t_2$ increases, since its trace is not conserved. The peak heights rapidly attenuate in the first few microseconds. Even so,  Fig.~\ref{fig:2DPlot} (h), with an extremely small colorbar (which is not shown here), is similar to Fig.~\ref{fig:2DPlot} (d). This indicates that, although the NHH method continues to lose its scale, it still preserves the relative information between peaks. Here, we have the second condition that should be satisfied when using the NHH method. This method is suitable for short-time dynamics and fails to observe the steady state in the long-time limit.

\begin{figure}[h]
\includegraphics[width=8.45 cm]{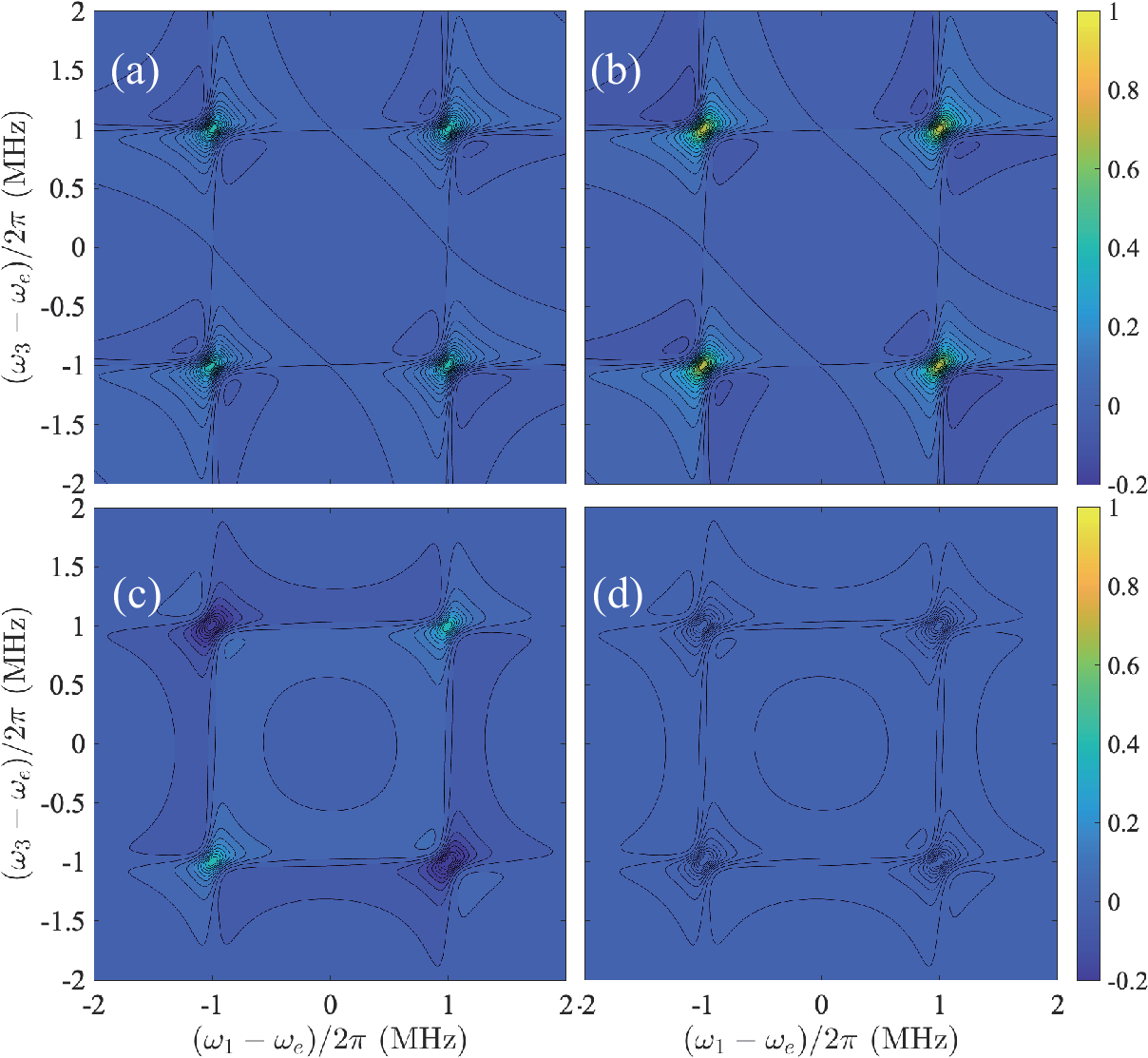}
\caption{Comparison of the NHH method's partial 2DCS with different evolving paths in $t_1$ and $t_3$. The first (second) row correspond to the summation of the paths evolving from $\vert b \rangle\langle e\vert$ to $\vert b \rangle\langle e\vert$ ($\vert b \rangle\langle e\vert$ to $\vert b \rangle\langle c\vert$) in $t_1$, while evolving from $\vert e \rangle\langle b\vert$ to $\vert e \rangle\langle b\vert$ ($\vert c \rangle\langle b\vert$ to $\vert e \rangle\langle b\vert$) in $t_3$. The 2DCS are plotted when (a,c) $t_2=0.24$~$\rm \mu s$, (b,d) $t_2=0.50$~$\rm \mu s$.}
\label{fig:2DPlot_divided}
\end{figure}

The four peaks of the RF formalism synchronously become lower at $t_2=~0.24~\rm \mu s$, then raise at $t_2=~0.5~\rm \mu s$. They exhibit synchronous oscillating behavior, which is dominated by $\mathcal{G}_{eeee}$. The NHH method, on the other hand, has little variance on the diagonal peaks despite attenuation, while the off-diagonal peaks continue to oscillate. For better clarity, the heights of the diagonal peak at (-1.005,-1.005)~MHz and the non-diagonal peak at (-1.005,0.995)~MHz are plotted as functions of $t_2$, as shown in Fig.~\ref{fig:slice} (a) and (b).

According to the Liouville paths of each term in $t_1$ and $t_3$, the RP signal of the NHH method is divided into three parts. The paths of Fig.~\ref{fig:FeynmanDiagram_NH}(a) and (b) are added together as the part that experiences $\vert b \rangle\langle e\vert$ evolving to $\vert b \rangle\langle e\vert$ during $t_1$, while $\vert e \rangle\langle b\vert$ evolving to $\vert e \rangle\langle b\vert$ during $t_3$. Figure~\ref{fig:FeynmanDiagram_NH}(c) and (d) are added together and have little effect on the final 2DCS. Figure~\ref{fig:FeynmanDiagram_NH}(e) is classified into a separate part for it experiences time evolution from $\vert b \rangle\langle e\vert$ to $\vert b \rangle\langle c\vert$ in $t_1$ and $\vert c \rangle\langle b\vert$ to $\vert e \rangle\langle b\vert$ in $t_3$.
The 2DCS of the first and the third parts are demonstrated in Fig.~\ref{fig:2DPlot_divided}. In Fig.~\ref{fig:2DPlot_divided}(c), the non-diagonal peaks are negative, which are caused by the emission peaks of the quasi-Green function in Fig.~\ref{fig:GreenFunction_w1w3}. The remaining positive peaks at $t_2=0.24$~$\rm \mu s$ are shallow, but are summed up to be the same scale in accordance with the peak heights at $t_2=0$~$\rm \mu s$. At $t_2=0.50$~$\rm \mu s$, on the other hand, Fig.~\ref{fig:2DPlot_divided}(d) gives little contribution and the final 2DCS is dominated by Fig.~\ref{fig:2DPlot_divided}(b). As a result, Fig.~\ref{fig:2DPlot_divided}(b) is almost the same as Fig.~\ref{fig:2DPlot}(g). These 2DCS are consistent with the population dynamics. In other words, having less population on the excited state $\vert e \rangle$ despite decay is the feature of stronger coherence between the energy levels $e$ and $c$, as well as more contribution from the Liouville path Fig.~\ref{fig:FeynmanDiagram_NH}(e).

Next, for a better comparison of the two approaches, we recalculate the 2DCS by the RF formalism  with Liouville paths the same as the NHH method in  Fig.~\ref{fig:FeynmanDiagram_NH}. The RP signal is written as
\begin{eqnarray} \label{Eq:RPsignal_resNHFun}
    S^{(3)}_{\rm RP}(\omega_{3},t_{2},\omega_{1})\!&=&\!\tilde{\mathcal{G}}_{ebeb}(\omega_{3})\mathcal{G}_{bbbb}(t_{2})\tilde{\mathcal{G}}_{bebe}(\omega_{1})\notag \\
    \!&&\!+\tilde{\mathcal{G}}_{ebeb}(\omega_{3})\mathcal{G}_{eeee}(t_{2})\tilde{\mathcal{G}}_{bebe}(\omega_{1})\notag \\
    \!&&\!+\tilde{\mathcal{G}}_{ebcb}(\omega_{3})\mathcal{G}_{ceee}(t_{2})\tilde{\mathcal{G}}_{bebe}(\omega_{1})\notag \\
    \!&&\!+\tilde{\mathcal{G}}_{ebeb}(\omega_{3})\mathcal{G}_{eeec}(t_{2})\tilde{\mathcal{G}}_{bcbe}(\omega_{1})\notag \\
    \!&&\!+\tilde{\mathcal{G}}_{ebcb}(\omega_{3})\mathcal{G}_{ceec}(t_{2})\tilde{\mathcal{G}}_{bcbe}(\omega_{1}).
\end{eqnarray}
Correspondingly, the additional Green functions are written as
\begin{align}
    \tilde{\mathcal{G}}_{bcbe}(\omega_{1})=&\frac{2\Omega_{ec}}{4(\omega_{1}-\omega_{cb}-i\gamma_{bc})(-\omega_{1}+\omega_{cb}+i\gamma_{eb})+\Omega_{ec}^{2}}, \\
    \tilde{\mathcal{G}}_{ebcb}(\omega_{3})=&\frac{2\Omega_{ec}}{4(\omega_{3}-\omega_{eb}+i\gamma_{bc})(-\omega_{3}+\omega_{eb}-i\gamma_{eb})+\Omega_{ec}^{2}}, \\
    \mathcal{G}_{ceee}(t_{2})=&\frac{\Omega_{ec}}{4\widetilde{\Omega}_{+}}\left(-1+e^{2i\widetilde{\Omega}_{+}t}\right)e^{-\frac{1}{2}(\gamma_{ec}^{+}+2i\widetilde{\Omega}_{+})t}, \\
    \mathcal{G}_{eeec}(t_{2})=&-\mathcal{G}_{ceee}(t_{2}), \\
    \mathcal{G}_{ceec}(t_{2})=&\frac{\gamma_{ec}^{+}\sin(\widetilde{\Omega}_{+}t)-2\widetilde{\Omega}_{+}\cos(\widetilde{\Omega}_{+}t)}{4\widetilde{\Omega}_{+}}e^{-\frac{1}{2}\gamma_{ec}^{+}t} \notag \\
    &+\frac{1}{2}e^{-\gamma_{ec}^{+}t}.
\end{align}
In this case, the 2DCS are shown in Fig.~\ref{fig:2DPlot} (i) to (l). Meanwhile, the variance of heights of the diagonal and non-diagonal peaks is plotted in Fig.~\ref{fig:slice} (c). It is shown that when the RF formalism and the NHH method are applied using the same Liouville paths, the resulting 2DCS exhibit quite similar behaviors. The only discrepancy  of the 2DCS occurs significantly in the long-time limit, where the peak heights by the NHH method markedly decrease as compared to the counterparts by the RF formalism. These results reveal that the two methods are mostly consistent when having the identical Liouville paths.

Considering that the Liouville paths of the RF formalism in Fig.~\ref{fig:FeynmanDiagram} do not include terms like $\vert b \rangle\langle c \vert$ or $\vert e \rangle\langle c \vert$ which are induced by the control field, we believe that the original RF formalism fails in accurately describing the time evolutions of the system and identifying all relevant Liouville paths. Conversely, the NHH method, enhanced by our simple improvement with the quasi-Green function, effectively provides the appropriate Liouville paths.

\begin{figure*}[h]
\includegraphics[width=16.5 cm]{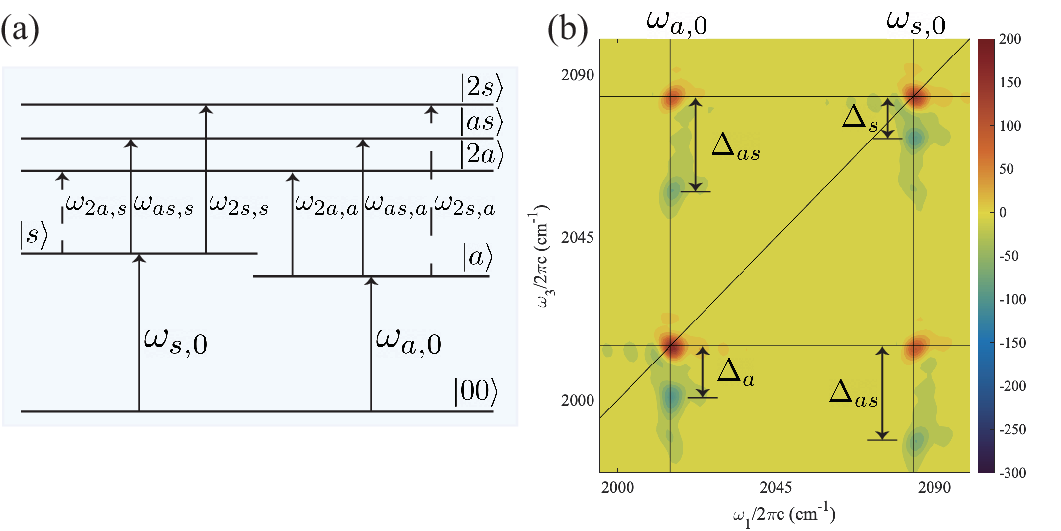}
\caption{(a) Schematic diagram of the six levels in RDC, which includes the ground vibrational state $\vert 00\rangle$, one-quantum states $\vert a\rangle$ and $\vert s\rangle$, as well as two-quantum states $\vert 2a \rangle$, $\vert 2s\rangle$ and $\vert as\rangle$. The transition frequencies are respectively $\omega_{a,0}=2015~\textrm{cm}^{-1}$, $\omega_{s,0}=2084~\textrm{cm}^{-1}$, $\omega_{2a,a}=2001~\textrm{cm}^{-1}$, $\omega_{2s,s}=2073~\textrm{cm}^{-1}$, $\omega_{as,a}=2058~\textrm{cm}^{-1}$, $\omega_{as,s}=1989~\textrm{cm}^{-1}$,
$\omega_{2a,s}=1932~\textrm{cm}^{-1}$, and $\omega_{2s,a}=2142~\textrm{cm}^{-1}$. \cite{Khalil2003JPCA,Demirdoven2002PRL}.
(b) 2DCS of RDC dissolved in hexane by the NHH method. 
The horizontal and vertical lines mark the frequencies $\omega_{a,0}$ and $\omega_{s,0}$. The gap values are approximately $\Delta_{as}=26~\textrm{cm}^{-1}$, $\Delta_{a}=14~\textrm{cm}^{-1}$, $\Delta_{s}=11~\textrm{cm}^{-1}$ \cite{Khalil2003JPCA,Demirdoven2002PRL}.
}
\label{fig:2DPlot_expr}
\end{figure*}

Finally, we simulate the 2DCS of RDC dissolved in hexane using the NHH method. The system has already been experimentally measured by Khalil et al. \cite{Khalil2003JPCA}. This system has six states, which include a ground vibrational state $\vert 00\rangle$, two one-quantum states, i.e., $\vert a\rangle$, $\vert s\rangle$, and three two-quantum states, i.e., $\vert 2a \rangle$, $\vert 2s\rangle$, $\vert as\rangle$. The spectral bandwidth of the three incident pulses is wide enough to excite all the vibrational transitions. The energy levels and transitions are shown in Fig.~\ref{fig:2DPlot_expr}(a), with dashed arrows representing forbidden transitions and solid arrows representing allowed transitions. The corresponding frequencies are as follows \cite{Khalil2003JPCA,Demirdoven2002PRL},  $\omega_{a,0}=2015~\textrm{cm}^{-1}$, $\omega_{s,0}=2084~\textrm{cm}^{-1}$, $\omega_{2a,a}=2001~\textrm{cm}^{-1}$, $\omega_{2s,s}=2073~\textrm{cm}^{-1}$, $\omega_{as,a}=2058~\textrm{cm}^{-1}$, $\omega_{as,s}=1989~\textrm{cm}^{-1}$,
$\omega_{2a,s}=1932~\textrm{cm}^{-1}$, and $\omega_{2s,a}=2142~\textrm{cm}^{-1}$, with the total decoherence rate $\gamma=0.3~\textrm{cm}^{-1}$. The elements of the electric dipole operator, describing the intensity of each transition, are given by $\mu_{a,0}=1.05\mu_{s,0},\mu_{as,s}=\mu_{a,0},\mu_{as,a}=\mu_{s,0},\mu_{aa,a}=1.48\mu_{s,0},\mu_{ss,s}=1.41\mu_{s,0},\mu_{ss,a}=\mu_{aa,s}=0.13\mu_{s,0}$ \cite{Golonzka2001}. Meanwhile, $t_1$ and $t_3$ vary from 0 to 5 ps, with a time step of $5~\rm{fs}$, and $t_2$ is set to 0. The pulse durations are neglected, as they contribute to the final results by an overall parameter.

The purely-absorptive spectrum is obtained by calculating the rephasing and non-rephasing signals. Zero padding is applied to expand the data length to 5000 points per dimension before performing the Fourier transform. After performing the double Fourier transforms, the real parts of the rephasing and non-rephasing signals are taken. The rephasing signal is then reflected against the $\it{y}$-axis and summed with the non-rephasing signal. As a result, the 2DCS for absorptive signals is shown in Fig.~\ref{fig:2DPlot_expr}(b).

Our simulation qualitatively reproduces the experimental features demonstrated by Khalil et al., including the number, positions, signs of the peaks in 2DCS \cite{Khalil2003JPCA}. The system can only be excited to the one-quantum states by the first pulse. As a result, the peaks appear along the $\omega_{1}$ axis only when $\omega_{1}=\omega_{a,0}, \omega_{s,0}$. Along the $\omega_{3}$ axis, on the other hand, there are 6 possible frequencies at which peaks appear. When $\omega_{3}=\omega_{a,0}, \omega_{s,0}$, formed by the excitation of the third pulse, the peaks have positive signs. When $\omega_{3}=\omega_{2a,a}, \omega_{2s,s}, \omega_{as,a}, \omega_{as,s}$, the peaks have negative signs indicating the radiations from the two-quantum states. The gaps between the positive peaks and the negative peaks are thus equal to the frequency differences between the radiations and fundamental transitions. These gaps, marked by double-headed arrows on the plot, are approximately $\Delta_{as}=26~\textrm{cm}^{-1}$, $\Delta_{a}=14~\textrm{cm}^{-1}$, $\Delta_{s}=11~\textrm{cm}^{-1}$, which are equal to the frequency differences \cite{Khalil2003JPCA}. Considering the well-fitted result, we believe the NHH method is a reliable theoretical approach for simulating the 2DCS.

\section{Conclusion} \label{sec:Conclusion}

In this paper, we compare the NHH method and the RF formalism with their advantages and disadvantages in the obtaining of 2DCS. In particular, we propose a simple way to simplify the computation processes of the NHH method by introducing the quasi-Green function, which improves its interpretive capacity. Both methods are demonstrated by gaining the RP signal of 2DCS for the same three-level system under the influence of a control field. The NHH signal is decomposed into the sum of the products, of which three quasi-Green functions corresponding to different $t_c~(c=1,2,3)$ are multiplied. An intuitive explanation of the NHH method is thus given directly in terms of the Feynman diagrams.
It is worth noting that the NHH method yields signal with more Liouville paths and evolution information. We compare the population dynamics of the two methods in $t_2$, obtaining similar oscillatory behavior in the first few periods, but different steady states for the long-time evolution. The comparison of the sum of populations on the three levels in $t_2$ indicates that for the RF formalism the probability is conserved, while for the NHH method the probability keeps decaying. This decay leads to the different steady states in the population dynamics, and also to a weakening of the emitted signal as $t_2$ increases. On this issue, the equation-of-motion method is capable of handling these scenarios despite its higher computational complexity \cite{Gelin2022CR}, i.e., $O(N^2)$ vs $O(N)$, with $N$ being the number of states. This method is powerful since it does not require the use of Feynman diagrams. Moreover, it can be combined with thermo field dynamics to account for more realistic temperature effects. \cite{Borrelli2016,Gelin-2017,Gelin-2021}

We plot the 2DCS for four representative $t_2$'s. Both methods yield splitting peaks with the same locations for each split peak. For the RF formalism, the heights of the four peaks oscillate synchronously. In contrast, the four peaks of the NHH method oscillate for a few periods with continuous attenuation, and the oscillation amplitudes of the non-diagonal peaks vary more significantly than those of the diagonal peaks. We show that the disappearance and emergence of off-diagonal peaks in 2DCS by the NHH approach are attributed to the different Feynman diagrams induced by the control field.  Furthermore, by comparing the 2DCS of the two methods with the same NHH Liouville paths, we show that the main difference between the two approaches is due to the introduction of some additional Feynman diagrams and the disappearance of some less-important Feynman diagrams induced by the control fields. The RF formalism in its original form is simply inapplicable to driven systems with time-dependent Hamiltonians. Though it can be applied if we switch to the rotating frame with a time-independent Hamiltonian. In addition to the simplified model system, we simulate the 2DCS of RDC dissolved in hexane. The positions and signs of all peaks are faithfully reproduced by the NHH approach, and is in good agreement with the experimental plot. Therefore, by comparing the 2DCS from experiments with those predicted by the NHH approach, we can effectively extract useful information about the structure and dynamics of the molecule under investigation.

Compared to the Hamiltonian in the RF formalism, the NHH includes the relaxation but drops the quantum-jump terms. Meanwhile, the RF formalism takes the relaxation into account by solving the quantum master equation without dropping these terms. This approximation in the NHH method leads to a non-conserved trace in population dynamics, and manifests as an overall decay in the intensity of the 2DCS. However, the overall decay does not significantly influence other phenomena, such as the oscillations in the off-diagonal peaks. Despite the overall decay, other differences arise from the method adopted, with the most significant being the disappearance and reappearance of off-diagonal peaks.

Overall, the main differences between the 2DCS of the two methods arise from the fact that, in the NHH method, the relaxations are equivalent to unidirectional population losses to the environment. Additionally, there are more significant Liouville paths contributing to the final 2DCS. The NHH method has its advantage of obtaining the signals directly from the analytical solution of time evolution, and thus would not omit any dominant Liouville paths. However, since we assume constant decay rates, the NHH method is not suited for lineshape analysis and thus can not account for homogeneous broadening. 
As for the RF formalism, its results are reliable as long as it considers all the important Liouville paths. 
To sum up, the NHH method can be more applicable in the systems with abundant energy structure and transition paths, as compared to the RF formalism. It will help to obtain 2DCS with all dominant Liouville paths considered and has prospects for exploring the dynamics of molecular or atomic systems with control fields.

\begin{acknowledgement}

We thank Lipeng Chen for valuable comments on the paper.
This work is supported by Innovation Program for Quantum Science and Technology under Grant No.~2023ZD0300200, the National Natural Science Foundation of China under Grant No.~62461160263, Beijing Natural Science Foundation under Grant No.~1202017, and Beijing Normal University under Grant No.~2022129.

\end{acknowledgement}

\bibliographystyle{apsrev4-2}
\bibliography{referenceVr}

\end{document}